\documentclass[prc,nofootinbib]{revtex4}
\usepackage{graphicx,bm,color}
\newcommand{\lead}{$^{209}{\rm Pb}$}

\newcommand{\mercury}{$^{199}{\rm Hg}$}
\newcommand{\radium}{$^{225}{\rm Ra}$}
\newcommand{\pb}{$^{208}{\rm Pb}$}

\newcommand{\hg}{$^{198}{\rm Hg}$}

\newcommand{\vv}{{a}}
\newcommand{\kk}{{k}}
\newcommand{\lll}{{l}}
\newcommand{\ii}{{i}}

\begin{document}
\bibliographystyle{apsrev}

\title{Time--Reversal--Violating Schiff Moment of \mercury}
\author{J. H. de Jesus}
\email[]{jhjesus@physics.unc.edu} \altaffiliation[New address:
]{Department of Physics, University of  Wisconsin, Madison, WI
53706}
\author{J. Engel}
\email[]{engelj@physics.unc.edu}
\affiliation{Department of Physics and Astronomy, CB3255, University of
North Carolina, Chapel Hill, NC 27599--3255}
\date{\today}
\begin{abstract}
We calculate the Schiff moment of the nucleus \mercury, created by
$\pi$NN vertices that are odd under parity (P) and time--reversal
(T).  Our approach, formulated in diagrammatic perturbation theory
with important core--polarization diagrams summed to all orders,
gives a close approximation to the expectation value of the Schiff
operator in the odd--A Hartree--Fock--Bogoliubov ground state
generated by a Skyrme interaction and a weak P-- and T--odd
pion--exchange potential.  To assess the uncertainty in the results,
we carry out the calculation with several Skyrme interactions, the
quality of which we test by checking predictions for the
isoscalar--E1 strength distribution in \pb, and estimate most of the
important diagrams we omit.
\end{abstract}
\pacs{24.80.+y,11.30.Er,21.60.Jz,27.90.+w}
\maketitle
%
%
\section{Introduction}
\label{sec:intro}

The existence of a permanent\footnote{Permanent or static, rather
than induced.} electric dipole moment (EDM) in leptons, neutrons or
neutral atoms is direct evidence for time--reversal (T) violation.
Because of the CPT theorem, the search for EDMs can provide us
valuable information about sources of CP violation.  Though a phase
in the Cabibbo--Kobayashi--Maskawa matrix is enough  to account the
level of CP violation in kaon and B--meson decays, it cannot explain
the observed matter/anti--matter asymmetry in the Universe.  Physics
that can (and new physics at the weak scale more generically) should
produce EDMs not far from current upper limits.

So far no EDMs have been observed, but experiments are continually
improving. Here we are interested in the conclusions that can be
drawn from the measured upper limit \cite{romalis01} in \mercury, a
diamagnetic atom. The largest part of its EDM
most likely comes from T violation in the nucleus, caused by a
T--violating (and parity--violating) component of the
nucleon--nucleon interaction. The atomic EDM is generated by the
subsequent interaction of the nucleus with the electrons.

That interaction is more subtle than one might think.  If the
nucleus and electrons were non--relativistic point--particles
interacting solely via electrostatic forces, the electrons would
rearrange in response to a nuclear EDM to cancel it essentially
exactly. Fortunately, as was shown by Schiff~\cite{schiff63}, the
finite size of the nucleus leads to a residual atomic EDM.  It turns
out, however, that the relevant nuclear quantity is not the nuclear
EDM but rather the nuclear ``Schiff moment''
\begin{equation} \label{eqn:sm}
S~\equiv~\langle\Psi_0|S^z|\Psi_0\rangle~, \end{equation}
which is the nuclear ground--state expectation value, in the
substate $|\Psi_0\rangle$ with angular momentum projection $M_J$
equal to the angular momentum $J$, of the $z$--component of the
``Schiff operator''
\begin{equation}
\label{eqn:op}
\bm{S}~=~\frac{e}{10}\,\sum_{p=1}^Z\left(r^2_p-\frac{5}{3}\,\langle{r}^2
\rangle_{\rm
ch}\right)\bm{r}_p~.
\end{equation}
Here $e$ is the charge of the proton, $\langle r^2\rangle_{\rm ch}$
is the mean squared radius of the nuclear charge distribution, and
the sum is restricted to protons.

For the Schiff moment to exist, P and T must be violated by the
nuclear Hamiltonian.  We assume that whatever its ultimate source,
the T violation works its way into a meson--mediated P-- and
T--violating NN interaction generated from a Feynman graph
containing a meson propagator, the usual strong meson--NN strong
vertex and a (much weaker) P-- and T--violating meson--NN vertex.
The second vertex can take three different forms in isospin space.
References~\cite{herczeg87, griffiths91, towner94} showed that
short--range nuclear correlations and a fortuitous sign make the
contribution of $\rho$-- and $\omega$--exchange to the interaction
small compared to that of pion--exchange if the T--violating
coupling constants of the different mesons are all about the same,
and so we neglect everything but pion exchange.  The most general
P-- and T--odd NN potential then has the form ($\hbar=c=1$)
\begin{eqnarray}
\label{eqn:wpt}
W(\bm{r}_a-\bm{r}_b)&=&-\frac{{\rm g}m_\pi^2}{8\pi m_{\rm N}}
\left\{\left[{\rm \bar{g}}_0\left(\bm{\tau}_a\cdot\bm{\tau}_b\right)
-\frac{{\rm \bar{g}}_1}{2}\left(\tau_a^z+\tau_b^z\right)+
{\rm \bar{g}}_2\left(3\tau_a^z\tau_b^z-\bm{\tau}_a\cdot\bm{\tau}_b\right)
\right]\left(\bm{\sigma}_a-\bm{\sigma}_b\right)-\right.\nonumber\\
&&\mbox{}\left.-\frac{{\rm \bar{g}}_1}{2}\left(\tau_a^z-\tau_b^z\right)
\left(\bm{\sigma}_a+\bm{\sigma}_b\right)\right\}\cdot
\left(\bm{r}_a-\bm{r}_b
\right)\frac{{\rm exp}\left(-m_\pi|\bm{r}_a-\bm{r}_b|\right)}
{m_\pi|\bm{r}_a-\bm{r}_b|^2}\left[1+\frac{1}{m_\pi|\bm{r}_a-\bm{r}_b|}
\right]~,
\end{eqnarray}
where $m_\pi$ is the mass of the pion, $m_{\rm N}$ that of the
nucleon, $\tau^z|p\rangle=-|p\rangle$, ${\rm g} \equiv 13.5$ is the
strong $\pi$NN coupling constant, and the $\bar{\rm g}_i$ are the
isoscalar ($i=0$), isovector ($i=1$), and isotensor ($i=2$) PT--odd
$\pi$NN coupling constants.  A word of caution here:  more than one
sign convention for the $\bar{\rm g}$'s is in use.  Our $\bar{\rm
g}_0$ and $\bar{\rm g}_1$ are defined with a sign opposite  to those
used by Flambaum et.\ al~\cite{flambaum86, flambaum02} and by
Dmitriev et.\ al~\cite{dmitriev03, dmitriev05}.

The goal of this paper is to calculate the dependence of the Schiff
moment of \mercury\ on the T--violating $\pi$NN couplings (we leave
the dependence of these couplings on fundamental sources of CP
violation to others) so that models of new physics can be
quantitatively constrained. An accurate calculation is not easy
because the Schiff moment depends on the interplay of the Schiff
operator with complicated spin-- and space--dependent correlations
induced by the the two--body interaction $W$.  In the early
calculation by Flambaum, Khriplovich and Sushkov almost two decades
ago~\cite{flambaum86}, the correlations were taken to be admixtures
of simple 1--particle 1--hole excitations into a Slater determinant
produced by a one--body Wood--Saxon potential.

More recent work~\cite{dmitriev03,dmitriev05} made significant
improvements by treating the correlations in the RPA after
generating an approximately self--consistent one--body potential.
However, that work used only the relatively schematic Landau--Migdal
interaction (in addition to $W$) in the RPA and mean--field
equations, and did not treat pairing self consistently. The reliance
on a single strong interaction makes it difficult to analyze
uncertainty. Such an analysis seems to be particularly important in
\mercury, the system with the best experimental limit on its EDM.
The calculated Schiff moment of Refs.~\cite{dmitriev03,dmitriev05}
in that nucleus depends extremely weakly on the isoscalar
coefficient ${\rm \bar{g}}_0$, a result of coincidentally precise
cancellations among single--particle and collective excitations.
They might be less precise when other interactions are used.

Here we make several further improvements.  Our mean field, which we
calculate in $^{198}$Hg before treating core polarization by the
valence nucleon, includes pairing and is exactly self consistent.
Pairing changes the RPA to the quasiparticle--RPA (QRPA), a
continuum version of which we use to obtain ground--state
correlations.  Most importantly, we carry out the calculation with
several sophisticated (though still phenomenological) Skyrme NN
interactions, the appropriateness of which we explore by examining
their ability to reproduce measured isoscalar--E1 strength
(generated by the isoscalar component of the Schiff operator) in
$^{208}$Pb.  The use and calibration of more than one such force
allows us to get a handle on the uncertainty in our final results.

The rest of this paper is organized as follows:  Section II
describes our approach and the Skyrme interactions we use, and
includes their predictions for strength distributions that bear on
the Schiff moment.  Section III presents our results and an analysis
of their uncertainty, including a calculation in the simpler nucleus
$^{209}$Pb that allows us to check the size of effects we omit in
\mercury.  Section IV is a conclusion.

\section{Procedure for evaluating Schiff moments}
\label{sec:method}
\subsection{Method}

Our Schiff moment is a close approximation to the expectation value
of the Schiff operator in the completely self--consistent
one--quasiparticle ground state of \mercury, constructed from a
two--body interaction that includes both a Skyrme potential and the
P-- and T--violating potential $W$.  It is an approximation because
we do not treat $W$ in a completely self consistent way, causing an
error that we estimate to be small in the Section \ref{sec:results}.
In addition, we do not actually carry out the mean--field
calculation in \mercury\ itself.  Instead, we start from the HF+BCS
ground--state of the even--even nucleus \hg\ and add a neutron in
the $2p_{1/2}$ level. We then treat the core--polarizing effects of
this neutron in the QRPA.  A self--consistent core with QRPA core
polarization is completely equivalent to a fully self--consistent
odd--A calculation~\cite{brown70}.  We omit one part of the QRPA
core polarization, again with an estimate showing its contribution
to be insignificant.

A good way to keep track of the two interactions and their effects
is to formulate the calculation (and corrections to it) as a sum of
Goldstone--like diagrams, following the shell--model
effective--operator formalism presented, e.g., in Ref.\
\cite{ellis97}.  The one difference between our diagrams and he
usual ``Brandow'' kind is that our fermion lines will represent BCS
quasiparticles rather than pure particles or holes.  Our diagrams
reduce to the familiar kind in the absence of pairing.

We begin, following a spherical HF+BCS calculation in \hg\ (in a 20--fm
box with mixed volume and surface pairing fixed as in Ref.\ \cite{bender02}),
by dividing
the Hamiltonian into unperturbed and residual parts. The
unperturbed part, expressed in the quasiparticle basis, is
\begin{equation}
\label{eqn:H_0}
H_0 = T + V_{00} + V_{11} ~,
\end{equation}
where $T$ is the kinetic energy and $V$ the Skyrme interaction, with
subscripts that refer to the numbers of quasiparticles the operator
creates and destroys. The residual piece\footnote{The BCS
transformation makes $V_{02}$ and $V_{20}$ zero.} is
\begin{equation}
\label{eqn:H_res}
H_{\rm res} = W + V_{22} + V_{13}+V_{31}+V_{04}+V_{40}~.
\end{equation}
The interaction $W$ can also be expanded in terms of quasiparticle
creation and annihilation operators; all the terms are included in
$H_{\rm res}$, though $W_{00}$ vanishes because $W$ is a
pseudoscalar operator. The  valence ``model space'' of
effective--operator theory is one--dimensional: a quasiparticle with
$u=0$ and $v=1$ (i.e.\ a particle, since it is not part of a pair)
in the the $a\equiv (2p_{1/2},m=1/2)$ level. The unperturbed ground
state $|\Phi_a\rangle$ is simply this one--quasiparticle state. Then
the expectation value of $S^z$, Eq.~(\ref{eqn:sm}), in the full
correlated ground state $|\Psi_{a}\rangle \equiv |\Psi_0\rangle$ is
given by
\begin{equation}
\langle\Psi_\vv|S^z|\Psi_\vv\rangle~=~{\cal N}^{-1}
\langle\Phi_\vv|\left[1+H_{\rm
res}\left(\frac{Q}{\epsilon_\vv-H_0}\right)+\cdots\right]S^z
\left[1+\left(
\frac{Q}{\epsilon_\vv-H_0}\right)H_{\rm
res}+\cdots\right]|\Phi_\vv\rangle~.
\label{eqn:per4}
\end{equation}
Here $\epsilon_\vv$ is the single--quasiparticle energy of the
valence nucleon, the operator $Q$ projects onto all other
single--quasiparticle states, ${\cal N}$ is a normalization factor
that, we will argue later, is very close to one, and the dots
represent higher--order terms in $H_{\rm res}$. To evaluate the
expression, we write $S^z$ in the quasiparticle basis as
$S^z=S_{11}+S_{02} + S_{20}$ ($S_{00}$ vanishes for the same reason
as $W_{00}$).

\begin{figure}[t]
\includegraphics{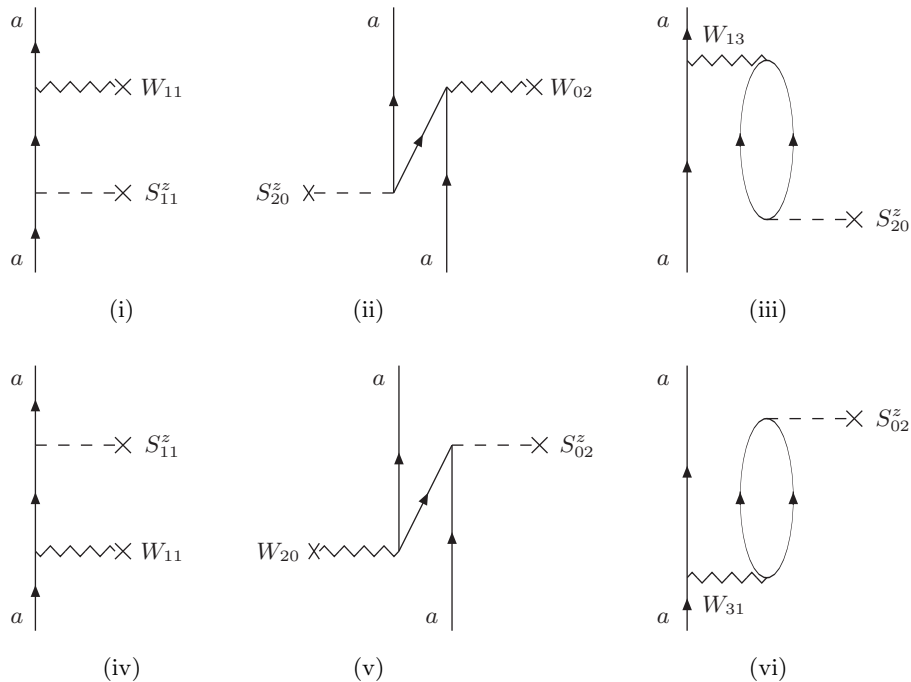}
\caption{First--order quasiparticle diagrams contributing to the
Schiff moment.  Diagrams (i), (ii), (iv) and (v) do not contribute
if the valence nucleon is a neutron, which is the case in \mercury.
Diagram (vi) is the complex conjugate of diagram (iii).}
\label{fig:order1}
\end{figure}

The zeroth--order contribution to the Schiff moment in Eq.\
(\ref{eqn:per4}) vanishes  because the Schiff operator cannot
connect two states with the same parity, and also because the Schiff
operator acts only on protons while the valence particle in
\mercury\ is a neutron. (There are no center--of--mass corrections
to the effective charges \cite{sushkov84}.)   The terms that are
first order in $H_{\rm res}$ do not include the strong interaction
$V$ because it has a different parity from the Schiff operator. Thus
the lowest order contribution to the Schiff moment is
\begin{equation}
\langle\Psi_\vv|S^z|\Psi_\vv\rangle^{\rm
first-order}~=~\langle-|q_\vv\left[W\left(\frac{Q}{\epsilon_\vv-H_0}\right)
S^z\right]
q^{\dag}_\vv|-\rangle+{\rm c.c.}~, \label{eqn:order1}
\end{equation}
where $q^{\dag}_\vv$ is the creation operator for a quasiparticle in
the valence level $\vv$ and $|-\rangle$ is the no--quasiparticle BCS
vacuum describing the even--even core, so that $|\Phi_a\rangle$ is
just $q^{\dag}_a|-\rangle$.  The contribution of
Eq.~(\ref{eqn:order1}) in an arbitrary nucleus can be represented as
the sum of the diagrams in Fig.~\ref{fig:order1}, the rules for
which we give in the Appendix\footnote{These rules are similar to
the ordinary Brandow/Goldstone--diagram rules but since fermion
lines represent quasiparticles, their number need not be conserved
at each interaction.  In addition, we have only upward going lines
because there are no quasiholes.}.  In \mercury, because the valence
particle is a neutron, only diagrams (iii) and (vi) are nonzero.  We
can interpret diagram (iii) as the Schiff operator exciting the core
to create a virtual three--quasiparticle state, which is then
de--excited back to the ground state when the valence neutron
interacts with the core through $W$.  This diagram and its partner
(vi) are what was evaluated by Flambaum et al.~\cite{flambaum86},
though their mean field was a simple Wood--Saxon potential, their
$W$ was a zero--range approximation that didn't include exchange
terms, and they neglected pairing.

\begin{figure}
\includegraphics{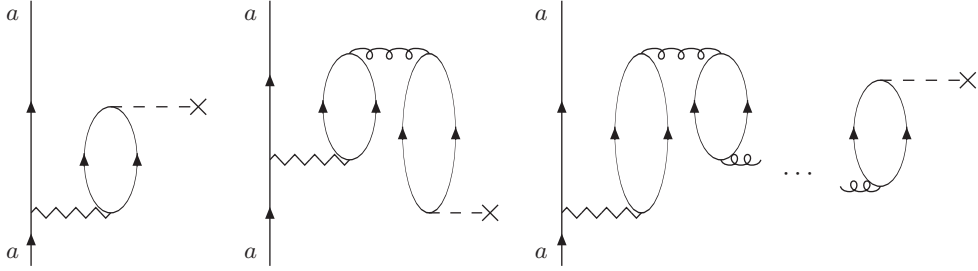}
\caption{Examples of diagrams contributing to the collective
response to the Schiff operator.  The sum of all these diagrams is
represented in Fig.~\ref{fig:orderRPA}(i).  The broken line
represents the action of the Schiff operator (as in Fig.\
\ref{fig:order1}), the zig-zag line represents the P-- and
T--violating interaction (also as in Fig.\ \ref{fig:order1}), and
the looped line represents a generic Skyrme interaction.}
\label{fig:QRPA}
\end{figure}

\begin{figure}[b]
\includegraphics{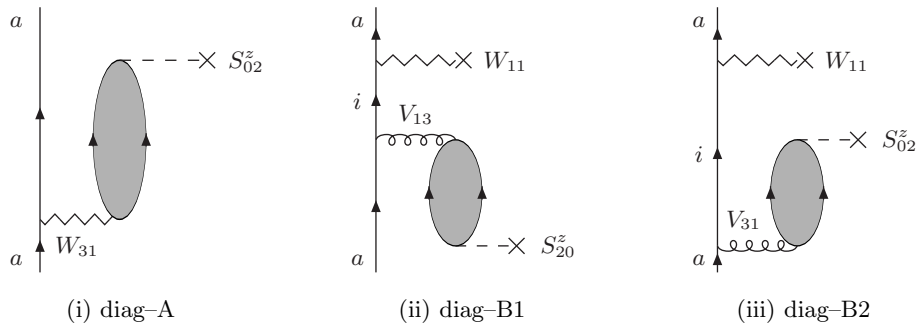}
\caption{QRPA diagrams contributing to the Schiff moment.  The
filled bubble represents an infinite sum of quasiparticle bubbles,
including all the forward and backward amplitudes.  The two B
diagrams have partners (not shown) in which $W$ acts below $S^z$.}
\label{fig:orderRPA}
\end{figure}

Core polarization, implemented through a version of the
canonical--basis QRPA code reported in Ref.\ \cite{terasaki05} (with
residual spurious center--of--mass motion removed following Ref.\
\cite{agrawal03}), can be represented by a subset of the
higher--order diagrams.  Because $W$ is so weak, we need only
include it in first order.  The higher order terms in $V$ that we
include have the effect of replacing the two--quasiparticle bubble
in (iii) and (vi) of Fig.~\ref{fig:order1} with chains of such
bubbles (see Fig.~\ref{fig:QRPA}), as well adding diagrams in which
the QRPA bubble chains are excited through a strong interaction of
the core with the valence neutron.  We therefore end up evaluating
diagrams labeled A, B1, and B2, in Fig.~\ref{fig:orderRPA} (plus two
more of type B in which the interaction $W$ is below the Schiff
operator).  The explicit expression for diagram A, the first on the
left in the figure, is
\begin{equation}
\label{eqn:diagA}
\langle\Psi_\vv|S^z|\Psi_\vv\rangle_{\rm diag-A}~=~-\sum_\lambda
\sum_{\kk>\lll}\sum_{k^{\prime}>l^{\prime}}Z^{\lambda\ast}_{\kk\lll}
\langle -|S^z_{02}|\kk\lll\rangle
Z^\lambda_{k^{\prime}l^{\prime}}\langle \vv k^{\prime}
l^{\prime} |W_{31}|\vv\rangle
{\cal E}_\lambda^{-1}~.
\end{equation}
Here, $Z^\lambda_{\kk\lll}\equiv
X^\lambda_{\kk\lll}+Y^\lambda_{\kk\lll}$ represents the QRPA
amplitudes (the sum appears because the matrix elements of all our operators
are real) and ${\cal E}_\lambda$
is the energy of the collective state $\lambda$ in \hg.
The quasiparticle matrix elements $\langle
-|S^z_{02}|\kk\lll\rangle$ and $\langle
\vv\kk\lll|W_{31}|\vv\rangle$ are related to the usual particle
matrix elements $\langle \kk|S^z|\lll\rangle$ and $\langle
\vv\kk|W|\vv\lll\rangle$ through the transformations discussed in
the Appendix.   In the absence of QRPA correlations, the $X$ and $Y$
amplitudes are 1 or 0, and Eq.~(\ref{eqn:diagA}) reduces to that
associated with diagrams (iii) and (vi) of Fig.~\ref{fig:order1}.

Diagrams B1 and B2 of Fig.~\ref{fig:orderRPA} have the
explicit expressions
\begin{eqnarray}
\langle\Psi_\vv|S^z|\Psi_\vv\rangle_{\rm diag-B1}&=&
-2\sum_\lambda\sum_\ii\sum_{\kk>\lll}\sum_{k^{\prime}>l^{\prime}}
\langle \vv|W_{11}|\ii\rangle Z^{\lambda\ast}_{\kk\lll}
\langle \ii|V_{13}|\vv\kk\lll\rangle Z^\lambda_{k{\prime}l^{\prime}}
\langle k^{\prime}l^{\prime}|S^z_{20}|-
\rangle(\epsilon_\vv-\epsilon_\ii)^{-1}
{\cal E}_\lambda^{-1}~,
\label{eqn:diagB1}\\
\langle\Psi_\vv|S^z|\Psi_\vv\rangle_{\rm diag-B2}
&=&-2\sum_\lambda\sum_\ii\sum_{\kk>\lll}\sum_{k^{\prime}>l^{\prime}}\langle
\vv|W_{11}|\ii\rangle Z^{\lambda\ast}_{\kk\lll}\langle
-|S^z_{02}|\kk\lll\rangle Z^\lambda_{k^{\prime}l^{\prime}}\langle
\ii k^{\prime}l^{\prime}|V_{31}|\vv\rangle
(\epsilon_\vv-\epsilon_\ii)^{-1}({\cal E}_\lambda
-\epsilon_\vv+\epsilon_\ii)^{-1}~.
\label{eqn:diagB2}
\end{eqnarray}
The factor 2 accounts for diagrams not shown in Fig.\
\ref{fig:orderRPA} in which $W$ acts {\it below} the QRPA bubble, and
the $\epsilon$'s are quasiparticle energies.  The difference between
Eq.~(\ref{eqn:diagB1}) and Eq.~(\ref{eqn:diagB2}) is mainly in 
the three-quasiparticle intermediate states.

A complete QRPA calculation that is first order in $W$ would also
include versions of diagram A in which $W$ trades places with one of
the $V$'s in the bubble sum.  We don't evaluate such diagrams but
estimate their size (which we find to be small)  rom calculations in
the simpler nucleus \lead\ in the next section.  We also use that
nucleus to examine other low--order diagrams not included in the
bubble sum of diagram A.

Why do we expect the QRPA subset of diagrams to be sufficient?  The
reason is that they generally account for the collectivity of
virtual excitations in a reliable way when calculated with Skyrme
interactions.  We illustrate this statement below with some
calculations of isoscalar--E1 strength in \pb.

\subsection{Interactions}

We carry out the calculation with 5 different Skyrme interactions:
our preferred interaction SkO$^{\prime}$ \cite{bender02,reinhard99}
(preferred for reasons discussed in Ref.\ \cite{engel03}), and the
older, commonly used interactions SIII \cite{beiner75}, SkM$^*$
\cite{bartel82}, SLy4 \cite{chabanat98}, and SkP
\cite{dobaczewski84}.  To get some idea of how well they will work,
we calculate the strength distribution of the isoscalar--E1 operator
\begin{equation}
\label{eqn:isgdr}
\bm{D}_0~=\sum_{p=1}^Zr_p^2\bm{r}_p+
\sum_{n=1}^Nr_n^2\bm{r}_n~.
\end{equation}
This operator is interesting because it is the isoscalar version of
the Schiff operator (the isoscalar version of the second term in the
Schiff operator acts only on the center of mass and so doesn't
appear in $\bm{D}_0$). The isoscalar--E1 strength, measured, e.g.,
in \pb\ \cite{clark01}, seems to fall mainly into two peaks. The
high--energy peak, related to the compressibility coefficient ${\rm
K}_\infty$,
\cite{davis97,hamamoto98,colo00,vretenar00,clark01,abrosimov02,
shlomo02}, is observed to lie between 19 and 23 MeV, depending on
the experimental method used \cite{davis97,clark01}.  Recent
interest has focused on a smaller but still substantial low--energy
peak around $12~{\rm MeV}$, which has been studied theoretically in
the RPA \cite{colo00,vretenar00} as well as experimentally
\cite{clark01}.

\begin{figure}[t]
\includegraphics[width=8cm,angle=-90]{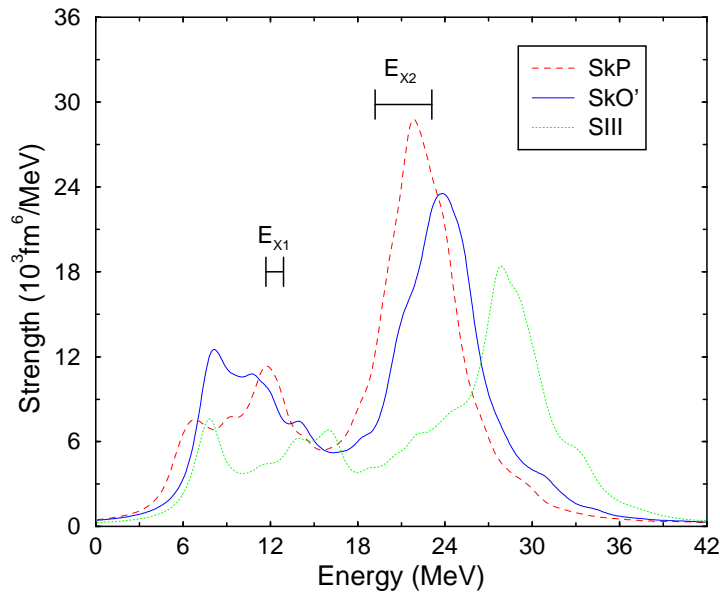}
\caption{Isoscalar--E1 strength distributions for \pb\ predicted by
the Skyrme interactions SkP, SkO$^\prime$ and SIII in
self--consistent HF+RPA. The experimental bounds on the low--energy
($\rm{E_{x1}}$) and high--energy (${\rm E_{x2}}$) peaks are also
shown (see Table~\ref{tb:isgdrpb}).} \label{fig:isgdrpb}
\end{figure}

Figure \ref{fig:isgdrpb} shows the predictions of several Skyrme
interactions in the RPA, with widths of 1 MeV introduced by hand
following Ref.\ \cite{hamamoto98}, for the isoscalar--E1 strength
distribution in \pb. The figure also shows the locations of the
measured low-- and high--energy peaks of Davis et al. \cite{davis97}
and of Clark et al. \cite{clark01}.  Nearly all self--consistent RPA
calculations, including ours (except with SkP) over--predict the
energy of the larger peak by a few MeV \cite{colo00, clark01}.  SIII
does a particularly poor job.  The predicted low--energy strength is
closer to experiment, though usually a little too low. Table
\ref{tb:isgdrpb} summarizes the situation.  Unfortunately, the data
are not precise enough to extract much more than the centroids of
the two peaks.  Since the Schiff--strength distribution in \mercury\
helps determine the Schiff moment, it would clearly be useful to
have better data, either in that nucleus or a nearby one such as
\pb.

The isovector--E1 strength distribution also bears on the Schiff
moment through the second term of the Schiff operator, but it is
well understood experimentally and generally reproduced fairly well
by Skyrme interactions.

\begin{table}[]
\begin{ruledtabular}
\begin{tabular}{lccc}
                               & & low (MeV) & high (MeV) \\
\hline
SkM$^\star$                    & & $11.0$ & $25.3$ \\
SkP                            & & $10.0$ & $23.4$ \\
SIII                           & & $11.6$ & $28.3$ \\
SLy4                           & & $11.4$ & $26.4$ \\
SkO$^\prime$                    & & $10.3$ & $24.8$ \\
\hline
Experiment~\cite{davis97}    & & $-$ & $22.4\pm0.5$ \\
Experiment~\cite{clark01}    & & $12.2\pm0.6$ & $19.9\pm0.8$ \\
\end{tabular}
\end{ruledtabular}
\caption{Comparison between experimental and theoretical results for
the centroids of the low- and high-energy peaks in the distribution
of isoscalar E1 strength in \lead. The experimental results are from
Refs.\ \cite{clark01} and \cite{davis97}. (Ref.\ \cite{davis97}
identifies only the high-energy peak.) The theoretical distributions
are from self-consistent HF+RPA calculations with five different
Skyrme interactions.} \label{tb:isgdrpb}
\end{table}

\section{Results and estimate of uncertainty}
\label{sec:results}

\subsection{Results with several forces}

The Schiff moment can be written as
\begin{equation}
S=a_0 ~ {\rm g \bar{g}}_0 +a_1 ~ {\rm g \bar{g}}_1 + a_2 ~ {\rm g
\bar{g}}_2
~,
\end{equation}
where ${\rm \bar{g}}_i$ are the P-- and T--odd $\pi$NN coupling
constants and all the nuclear physics is summarized by the three
coefficients $a_i$. We present our results for these coefficients by
showing the effects, in turn, of several improvements on early
calculations.

The first calculations of Schiff moments~\cite{flambaum86}, as noted
above, correspond to our first--order diagrams (iii) and (vi) of
Fig.\ \ref{fig:order1} but with no pairing, with a simple
Wood--Saxon potential in place of a self--consistent mean field, and
with the zero--range limit of the direct part of $W$.  The results
of Ref.\ \cite{flambaum86} are given here in the first line of Tab.\
\ref{tb:result}.

\begin{table}[t]
\begin{ruledtabular}
\begin{tabular}{lccc}
                               &$a_0$& $a_1$ & $a_2$ \\
\hline
Ref.\ \cite{flambaum86}        & $0.087$ & $0.087$ & $0.174$ \\
Naive limit                    & $0.095$ & $0.095$ & $0.190$ \\
Diagram A only                 & $0.018$ & $0.034$ & $0.031$ \\
\hline
Full result                    & $0.010$ & $0.074$ & $0.018$ \\
\end{tabular}
\end{ruledtabular}
\caption{Calculated coefficients $a_i$ from Ref.\ \cite{flambaum86}
and with the Skyrme interaction SkO$^{\prime}$ in several limits
(see text).  The full result is in the last line.} \label{tb:result}
\end{table}
When we repeat the calculation, evaluating diagrams (iii) and (vi)
with $W$ approximated by its direct part in the zero--range limit and
with the mean field from the Skyrme interaction SkO$^\prime$ (so
that the only differences in the calculations are the one--body
potential and BCS paring), we get the coefficients in second line of the
table.

The finite range of the potential reduces the $a_i$ from these
zero--range values by 30--40\%, depending on the Skyrme interaction
used.  Exchange terms, when the range is finite, decrease $a_0$ by a
few percent, have no effect on $a_1$ and increase $a_2$ by half the
amount they decrease $a_0$.

The three coefficients in lines 1 and 2 of the table are not
independent; the isotensor coefficient is exactly two times larger
than the isovector and isoscalar coefficients.  Because the valence
neutron must excite core protons to couple to the Schiff operator
through (iii) and (vi) of Fig.~\ref{fig:order1}, only the
neutron--proton part of $W$ contributes, and under the assumptions
of no core spin, no exchange terms, and zero--range, $W$ reduces to
\begin{equation}
W^{\rm contact}_{\rm direct}(\bm{r}_{n}-\bm{r}_p)~=~-\frac{1}
{2m_\pi^2m_{\rm
N}}({\rm g}{\rm \bar{g}}_0+{\rm g}{\rm \bar{g}}_1+2{\rm g}{\rm
\bar{g}}_2)\bm{\sigma}_n\cdot\bm{\nabla}_n\delta(\bm{r}_{n}-
\bm{r}_p)~,
\qquad {\rm no~core~spin}
\label{eqn:4wpt}
\end{equation}
Thus, in the approach of Flambaum et al.~\cite{flambaum86}, the
Schiff moment is a function of a single parameter, usually called
$\eta_{np}$. Exchange terms add another independent parameter and
QRPA bubbles bring in a third.  This last term arises because the
valence neutron, besides exciting a proton quasiparticle pair in the
core, can now excite neutron quasiparticle pairs that annihilate and
create proton pairs inside the bubble that then couple to the Schiff
operator (see Fig.~\ref{fig:QRPA}). Thus, in our complete
calculation, particularly when the B diagrams are included, the
three coefficients $a_i$ are independent.

\begin{figure}[b]
\includegraphics[width=9cm,angle=-90]{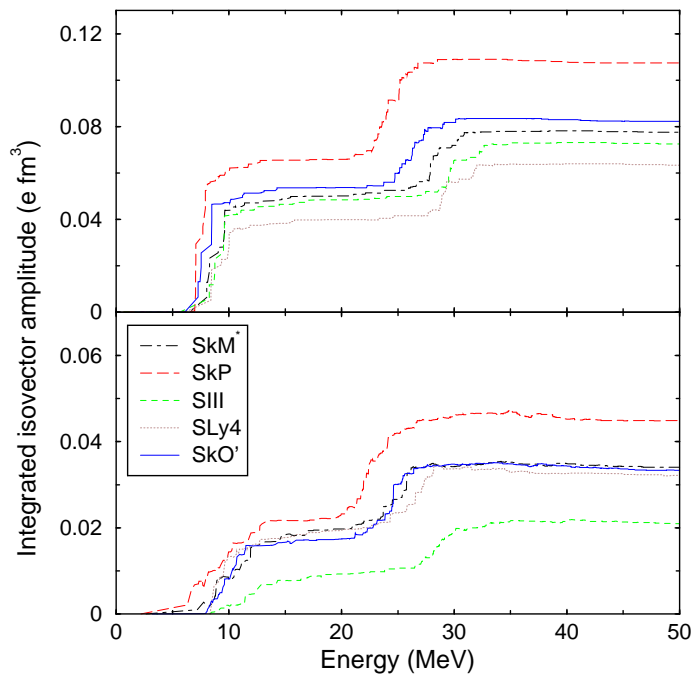}
\caption{Integrated contribution of ${\rm \bar{g}}_1$ to diagram A
as a function of core--excitation energy  (the value of the
isovector part of the diagram is given by the lines at large energy)
in the two--quasiparticle limit (top) and in the QRPA (bottom) for
five different Skyrme interactions.   The two bumps are at the
energies of the low-- and high--lying peaks of the isoscalar--E1
strength distribution.} \label{fig:civ}
\end{figure}

The collectivity of the core turns out to be very important.  As can
be seen in the third line of Tab.\ \ref{tb:result}, when the
single--particle bubble in the calculation above is replaced by the
full QRPA bubble sum to give diagram A, all three $a_i$ shrink {\em
substantially}.  The reason is that the Schiff strength is pushed on
average  to higher energies, both in the low--lying and high--lying
analogs of the isoscalar--E1 distribution. (The high--lying peak
actually is replaced by two peaks, the higher of which is at about
38 MeV. There is no peak corresponding to the giant isovector--E1
resonance, as was shown in Ref.\ \cite{engel99a}.)  The reduction is
greater in the isoscalar and  tensor channels --- a factor of 4 to 6
depending on the Skyrme interaction --- than in the isovector
channel, where it is a factor of 2 to 3.  Figure \ref{fig:civ} shows
the integral of the contribution to diagram A as a function of
core--excitation energy (so that at large energy the lines approach
the value for the diagram) for $a_1$, with and without the bubble
sum, for all 5 forces.

\begin{figure}[h]
\includegraphics[width=9cm,angle=-90]{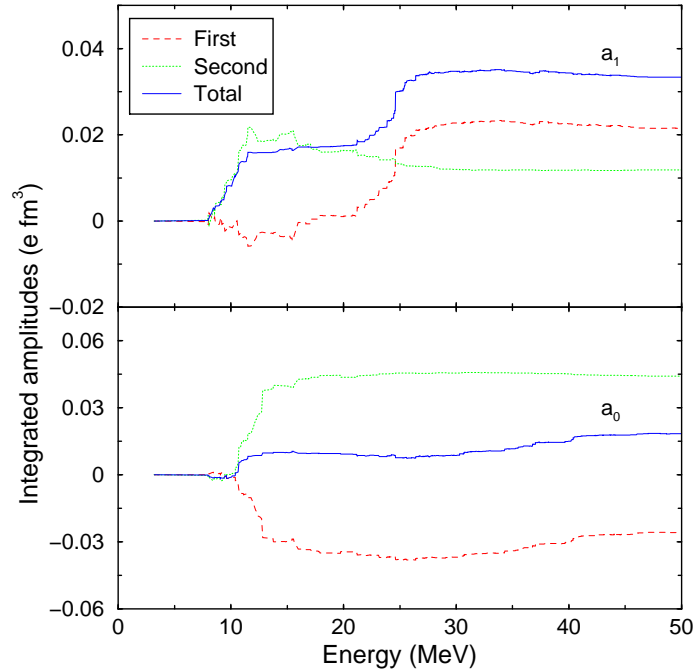}
\caption{Integrated contributions of the first (dashed) and second
(dotted) terms in the Schiff operator, together with the total
contribution (solid) to diagram A, as a function of core--excitation
energy with the SkO$^{\prime}$ interaction.  The top panel is for
the isovector coefficient $a_1$ and the bottom for the isoscalar
coefficient $a_0$.} \label{fig:parts}
\end{figure}

The reason for the difference in the size of the reduction is that
the ${\rm \bar{g}}_0$ and ${\rm \bar{g}}_2$ parts of the interaction
affect protons and neutrons in opposite ways (see, e.g., Eq.\ (9) of
Ref.\ \cite{engel03}), causing a destructive interference, while the
${\rm \bar{g}}_1$ part affects them in the same way. This difference
is absent from the single--quasiparticle picture because neutron
excitations of the core don't play a role there.  Another way of
saying the same thing is that when the neutrons and protons are
affected in the same way, the second (dipole--like) term in the
Schiff operator, Eq.\ (\ref{eqn:op}), contributes very little
because the center of mass and center of charge move together.  In
the other two channels the contribution of the second term is
similar in magnitude and opposite in sign to that of the first term
so that, as Fig.\ \ref{fig:parts} shows, the net value is smaller.

The type--B diagrams (see Fig.~\ref{fig:QRPA}) are important
corrections to diagram A.   The effective weak isoscalar and
isotensor one--body potentials (i.e.\ the tadpole) contribute with
opposite sign from that of the isovector potential; again, see Eq.\
(9) of \cite{engel03}, which displays the direct part of the
one--body potential explicitly.  The sign turns out to be opposite
that of diagram A in the isoscalar and isotensor channels, further
suppressing $a_0$ and $a_2$, and the same as diagram A in the
isovector channel, largely counteracting the suppression by
collectivity in that diagram.  The net result for SkO$^{\prime}$ is
in the last line of Tab.\ \ref{tb:result}; for the other forces the
net results appear in Tab.\ \ref{tb:skyrme}.  The isovector
coefficient $a_1$ ends up not much different from the early estimate
of Ref.\ \cite{flambaum86} but the isoscalar coefficient $a_0$ is
smaller by a factor of about 9 to 40 and the isotensor coefficient
$a_2$ by a factor of about 7 to 16.

\begin{table}[b]
\begin{ruledtabular}
\begin{tabular}{lccc}
                               &  $a_0$ & $a_1$  & $a_2$ \\
\hline
SkM$^\star$                    & $0.009$ & $0.070$ & $0.022$ \\
SkP                            & $0.002$ & $0.065$ & $0.011$ \\
SIII                           & $0.010$ & $0.057$ & $0.025$ \\
SLy4                           & $0.003$ & $0.090$ & $0.013$ \\
SkO$^\prime$                   & $0.010$ & $0.074$ & $0.018$ \\
\end{tabular}
\end{ruledtabular}
\caption{Full coefficients $a_i$ in \mercury\ for the five different Skyrme
interactions used here.  The units are $e~{\rm fm}^3$.} \label{tb:skyrme}
\end{table}

\subsection{Uncertainty and final result}

The several Skyrme interactions we use all give different results,
but the spread in numbers is about a factor of four in the isoscalar
channel, two in the isotensor channel, and much less for the large
isovector coefficient $a_1$.  It is possible that all the
interactions are systematically deficient, but we have no evidence
for that.  In any event, the effective interaction is not the only
source of uncertainty.  We have evaluated only a subset of all
diagrams, and although it is not obvious whether all the rest should
be evaluated with effective interactions that are determined through
Hartree--Fock-- or RPA--based fits, there are some that should
certainly be included and we would like to estimate their size.

\begin{figure}[h]
\includegraphics{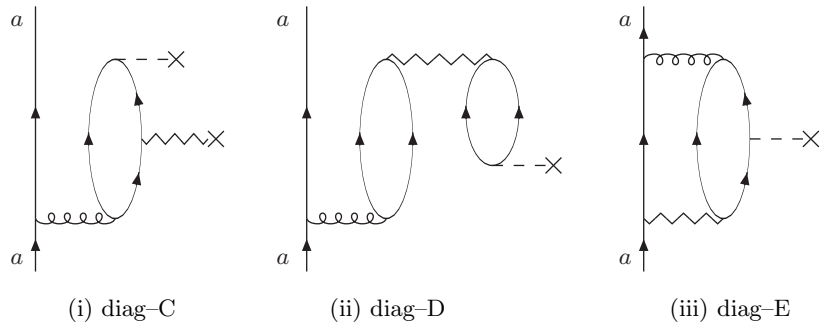}
\caption{Diagrams we did not include in our calculation in \mercury\
but the value of which we estimated through calculations in \lead\
(and \mercury\ in the case of diagram C).  We have omitted the
labels on the the interactions; they are the same as in the earlier
figures.}
\label{fig:others}
\end{figure}

The diagrams labeled C and D in Fig.\ \ref{fig:others} are the
leading terms in bubble chains that would result from including $W$
in the Hartree--Fock calculation (C) and in the QRPA calculation (D)
in \hg. We evaluated both sets in the simple nucleus \lead\ (which
has no pairing at the mean--field level), and found that diagram C
can be nearly as large as the type--B diagrams in the isovector and
isotensor channels. The same is true of diagram D in the isotensor
channel. In that nucleus, however, diagram A is much larger than all
the others and essentially determines the $a_i$.  In \mercury, we
only evaluated diagram C, but found that even though diagrams A and
B can cancel there, they do so the most in the isoscalar channel,
where the diagrams C and D are smallest.  In the end diagram C never
amounts to more than 10\% of the sum of the A and B diagrams, and
usually amounts to much less. Including the higher order (QRPA)
terms in the bubble chain will only reduce the diagram-C
contribution, so we conclude that it can be neglected.  We are not
positive that the same statement is true of diagram D, but unless it
is much larger in \mercury\ than in \lead\ (none of the other
diagrams are), it can be neglected too.

The diagram labeled E represents a correction from outside our
framework that is of the same order as the terms we include. We
evaluated it in \lead; it is uniformly smaller than those of type C
and D. Unless the situation is very different in \mercury, it can be
neglected as well.  The fact that these extra diagrams are all small
is not terribly surprising; they all bring in extra energy
denominators and/or interrupt the collective bubble.

We have also not included the normalization factor ${\cal N}$ in
Equation~(\ref{eqn:per4}). When calculated to second order in \lead,
it is about 1.05, independent of the Skyrme interaction used. Though
this factor could be larger in RPA order because of low--lying
phonons, most strength is pushed up by the RPA and we do not
anticipate a large increase.  It is reasonable to assume these
statements are true in \mercury\ as well.

At short distances the NN potential is strongly repulsive and the
associated short--range correlations should be taken into account.
Reference~\cite{griffiths91}, however, reports that the correlations
reduce matrix elements of the effective one--body P-- and
T--violating pion--exchange potential only by about 5\%, and in
Ref.\ \cite{dobaczewski05}, which calculates the Schiff moment of
\radium, their effects are smaller than 10\%. We are not missing
much by neglecting them, though we would be if we included a
$\rho$--meson exchange potential.

When all is said and done, the uncertainty is dominated by our
uncertainty in the effective interaction.  Our preferred interaction
is SkO$^{\prime}$, which (to repeat) gives the result
\begin{equation}
S_{^{199}{\rm Hg}}^{\rm SkO^{\prime}}~=~0.010{\rm g\bar{g}}_0+
0.074{\rm g\bar{g}}_1+
0.018{\rm g\bar{g}}_2~~~[e~{\rm fm}^3]~.
\label{eqn:myresults2}
\end{equation}
If instead we average the results from the five interactions, we get
\begin{equation}
S_{^{199}{\rm Hg}}^{\rm ave}~=~0.007{\rm g\bar{g}}_0+0.071{\rm g\bar{g}}_1+
0.018{\rm g\bar{g}}_2~~~[e~{\rm fm}^3]~,
\label{eqn:averesults}
\end{equation}
The range of results in Tab.\ \ref{tb:skyrme} is a measure of the uncertainty.

As noted in the introduction, Refs.\ \cite{dmitriev03,dmitriev05}
contain a similar calculation. They report
\begin{equation}
S_{^{199}{\rm Hg}}^{\rm Ref.\ [8]}~=~0.0004{\rm
g\bar{g}}_0+0.055{\rm g\bar{g}}_1+ 0.009{\rm g\bar{g}}_2~~~[e~{\rm
fm}^3]~,\label{eqn:hgdmitri}
\end{equation}
the most striking aspect of which is the isoscalar coefficient
$a_0$; it is more than an order of magnitude smaller than our
preferred value and five times smaller than the smallest coefficient
produced by any of our interactions.  We see no fundamental reason
for such serious suppression, and suspect that the same cancellation
we observe is coincidentally more precise in the single Hg
calculation of Refs.\ \cite{dmitriev03,dmitriev05}.  The authors
applied their method to other nuclei, but did not find the same
level of suppression in any of them.  Even the cancellation produced
in our calculations by SkP and SLy4 seems coincidentally severe.
Though it is possible that other realistic Skyrme interactions would
produce still smaller coefficients, we have a hard time imagining
it.

\section{Conclusions}
\label{sec:conclusions}

Our goal has been a good calculation of the dependence of the Schiff
moment of \mercury, the quantity that determines the electric dipole
moment of the corresponding atom, on three P-- and T--violating
$\pi$NN coupling constants. The current experimental limit on the
dipole moment of the \mercury\ atom, $|d| < 2.1\times 10^{-28}
e~{\rm cm}$, together with the theoretical results of Ref.\
\cite{dzuba02}, $d=-2.8 \times 10^{-17}(S/e~{\rm fm}^3)~e~{\rm cm}$,
yields the constraint $|S|<7.5\times 10^{-12} e~{\rm fm}^3$.  The
$a_i$ calculated in this paper then give a constraint on the three
${\rm \bar{g}}_i$.

In obtaining the $a_i$ we have included what we believe to be most
of the important physics, including a pion--exchange P-- and
T--violating interaction, collective effects that are known to
renormalize strength distributions of Schiff--like operators,
pairing at the mean--field level, self--consistency, and finally,
several different Skyrme interactions.  The last of these, together
with an examination of effects we omitted, allows us to give the
first real discussion of uncertainty for a calculation in this
experimentally important nucleus.

We conclude that while the isovector coefficient $a_1$ is not very
different from the initial estimate of Ref.\ \cite{flambaum86}, the
isoscalar coefficient, which determines the limit one can set on the
QCD parameter $\bar{\theta}$, is smaller by between about 9 and 40
(with the former our preferred value) and the isotensor parameter
$a_2$  by a factor between about 7 and 16 (with our preferred value
about 10). The uncertainty in these numbers comes primarily from our
lack of knowledge about the effective interaction.  There is good
reason to make better measurements of low--lying dipole strength,
particularly in the isoscalar channel.  They would help to unravel
the details of nuclear structure that determine the Schiff moment.

\begin{acknowledgments}
We thank J.\ Dobaczewski and J.\ Terasaki for helpful discussions.
This work was supported in part by the U.S.\ Department of Energy
under grant DE-FG02-97ER41019 and by the Funda\c c\~ao para a
Ci\^encia e a Tecnologia (Portugal).  J. H. de Jesus thanks the
Institute for Nuclear Theory at the University of Washington for its
hospitality and the Department of Energy for partial support during
the completion of this work.
\end{acknowledgments}

\section{Appendix}
\label{sec:appendix}

\subsection{Rules for quasiparticle diagrams in the uncoupled basis}

There are some differences between our rules for quasiparticle
diagrams and the usual rules for particle--hole diagrams.  The main
one is that one-- and two--body operators are written in a
quasiparticle basis and do not conserve quasiparticle number,
leading to different expressions for matrix elements. An example is
the generic quasiparticle operator $O_{20}$, which contains two
quasiparticle creation and no destruction operators. Its
matrix elements will be written $\langle kl|O_{20}|-\rangle$, which
means that it creates two quasiparticle states $|k\rangle|l\rangle$
out of the quasiparticle vacuum $|-\rangle$.

In what follows, ``in" refers to lines with arrows pointing toward
the vertex and ``out" to lines pointing away from it.  A diagram
should be read from  top to bottom, and from left to right. The
rules are then:
\begin{enumerate}
\item Each operator $O_{11}$ contributes
$\langle {\rm out}|O_{11}|{\rm in}\rangle$;
\item Each operator $O_{20}$ contributes
$\langle {\rm out,out^{\prime}}|O_{20}|-\rangle$;
because the diagram is read from the left, the label ``out" is on
the line that is further to the left;
\item Each operator $O_{02}$ contributes
$\langle-|O_{02}|{\rm in,in^{\prime}}\rangle$;
\item Each operator $O_{22}$ contributes
$\langle {\rm out,out^{\prime}}|O_{22}|
{\rm in,in^{\prime}}\rangle$;
\item Each operator $O_{31}$ contributes $\langle {\rm out,out^{\prime},
out^{\prime\prime}} |O_{31}|{\rm in}\rangle$;
\item Each operator $O_{13}$ contributes
$\langle {\rm out} |O_{13}|{\rm in,in^{\prime},
in^{\prime\prime}}\rangle$;
\item Each operator $O_{40}$ contributes
$\langle {\rm out,out^{\prime},out^{\prime\prime}
,out^{\prime\prime\prime}}|O_{40}|-\rangle$;
\item Each operator $O_{04}$ contributes
$\langle-|O_{04}|{\rm in,in^{\prime},
in^{\prime\prime},in^{\prime\prime\prime}}\rangle$;
\item The diagram should be summed over all intermediate states;
\item Energy denominators are evaluated by operating with
$(\epsilon_a-H_0)^{-1}$
between the action of every two operators in the diagram, giving
$[\epsilon_a+\sum_k\epsilon_k]^{-1}$, where $\epsilon_k$ are
quasiparticle energies;
\item The phase for each diagram is $(-)^{n_l}$, where $n_l$ is the
number of closed loops.
\item A factor of $1/2$ is included for each pair of
lines that start at the same vertex and end at the same vertex.
\end{enumerate}
Folded diagrams with additional rules occur in general, but we do
not discuss them here.

\subsection{Matrix elements of quasiparticles operators}

We first summarize some important quantities involving the
quasiparticle creation and annihilation operators $q^{\dag}$ and $q$,
which are defined in terms of the usual particle operators $a^{\dag}$ and
$a$ by
\begin{eqnarray}
\left\{\begin{array}{c}q_k=u_ka_k-v_k\tilde{a}_k^{\dag}\\
q_k^{\dag}=u_ka_k^{\dag}-v_k
\tilde{a}_k\end{array}\right .,\nonumber\\
\label{eqn:quasiparticles2}\\
\left\{\begin{array}{c}\tilde{q}_k=u_k\tilde{a}_k+v_ka_k^{\dag}\\
\tilde{q}_k^{\dag}
=u_k\tilde{a}_k^{\dag}+v_ka_k\end{array}\right ..\nonumber\\\nonumber
\end{eqnarray}
Here
\begin{equation}
\tilde{q}_k~\equiv~\tilde{q}_{l_kj_km_k}~=~(-)^{l_k+j_k+m_k}q_{l_kj_k-m_k}~=~
(-)^{l_k+j_k+m_k}q_{-k}~. \label{eqn:qpproperties1}
\end{equation}
From the anti--commutation rules for  $a^{\dag}$ and $a$, we derive the
following anti--commutation rules for the quasiparticle operators in
Eq.~(\ref{eqn:quasiparticles2})
\begin{eqnarray}
\begin{array}{c}
\label{eqn:qpproperties2}
\{q_k,q_l\}=\{q^{\dag}_k,q^{\dag}_l\}=\{\tilde{q}_k,
\tilde{q}_l\}=\{\tilde{q}^{\dag}_k,
\tilde{q}^{\dag}_l\}=\{\tilde{q}_k,q_l\}=\{\tilde{q}^{\dag}_k,q^{\dag}_l\}=0\\
\{q_k,q^{\dag}_l\}=\{\tilde{q}_k,\tilde{q}^{\dag}_l\}=\delta_{kl}\\
\{\tilde{q}_k,q^{\dag}_l\}=(-)^{l_k+j_k+m_k}\delta_{-kl}
\end{array}\\\nonumber
\end{eqnarray}

Using definition~(\ref{eqn:quasiparticles2}) and
properties~(\ref{eqn:qpproperties1}) and~(\ref{eqn:qpproperties2}),
one can  write a one--body operator in second quantization as
\begin{equation}
T=\sum_{kl}T_{kl}a^{\dag}_ka_l=T_0+T_{11}+T_{20}+T_{02}~,
\label{eqn:qpoperator1}
\end{equation}
where
\begin{eqnarray}
T_0&=&\sum_kv^2_kT_{kk}~, \\
T_{11}&=&\sum_{kl}T_{kl}(u_ku_lq^{\dag}_kq_l-v_kv_l
\tilde{q}^{\dag}_l\tilde{q}_k)~,\\
T_{20}&=&\sum_{kl}T_{kl}u_kv_lq^{\dag}_k\tilde{q}^{\dag}_l~,
\label{eqn:t20}\\
T_{02}&=&\sum_{kl}T_{kl}u_lv_k\tilde{q}_kq_l~.\\\nonumber
\label{eqn:qpoperator3}
\end{eqnarray}
Here, the subscripts indicate the number of quasiparticle creation
and annihilation operators involved. In the same way, a two--body
operator takes the form
\begin{eqnarray}
V&=&-\frac{1}{4}\sum_{klij}(V_{klij}-V_{klji})a^{\dag}_ka^{\dag}_la_ia_j=-
\frac{1}{4}\sum_{klij}
\overline{V}_{klij}a^{\dag}_ka^{\dag}_la_ia_j=\nonumber\\
&=&V_0+V_{11}+V_{20}+V_{02}+V_{22}+V_{31}+V_{13}+V_{40}+V_{04}~,
\label{eqn:qpoperator4}
\end{eqnarray}
where
\begin{eqnarray}
V_0&=&\frac{1}{4}\sum_{kl}\left(u_kv_ku_lv_lP_kP_l
\overline{V}_{k-kl-l}+2v_k^2v_l^2
\overline{V}_{klkl}\right)~, \\
V_{11}&=&\frac{1}{2}\sum_{kli}u_kv_lu_iv_iP_i
\overline{V}_{kli-i}\left(q^{\dag}_k
\tilde{q}_l+\tilde{q}^{\dag}_lq_k\right)\nonumber\\
&&+\mbox{}\sum_{kli}v_k^2\overline{V}_{klki}
\left(u_lu_iq^{\dag}_lq_i-v_lv_i\tilde{q}^{\dag}_i
\tilde{q}_l\right)~,\\
V_{20}&=&\frac{1}{4}\sum_{kli}u_iv_iP_i
\overline{V}_{kli-i}\left(u_ku_lq^{\dag}_kq^{\dag}_l+v_kv_l
\tilde{q}^{\dag}_l\tilde{q}^{\dag}_k\right)\nonumber\\
&&+\mbox{}\sum_{kli}u_kv_l^2v_i
\overline{V}_{klil}q^{\dag}_k\tilde{q}^{\dag}_i,\\
V_{22}&=&\frac{1}{4}\sum_{klij}\overline{V}_{klij}
\left(u_ku_lu_iu_jq^{\dag}_kq^{\dag}_lq_jq_i+v_kv_lv_iv_j
\tilde{q}^{\dag}_j\tilde{q}^{\dag}_i
\tilde{q}_k\tilde{q}_l\right)\nonumber\\
&&+\mbox{}\sum_{klij}u_kv_lu_jv_i\overline{V}_{klij}q^{\dag}_k
\tilde{q}^{\dag}_i
\tilde{q}_lq_j~,\\
V_{31}&=&\frac{1}{2}\sum_{klij}u_kv_l\overline{V}_{klij}
\left(u_iu_jq^{\dag}_iq^{\dag}_j
\tilde{q}^{\dag}_lq_k+v_iv_jq^{\dag}_k\tilde{q}^{\dag}_j\tilde{q}^{\dag}_i
\tilde{q}_l\right)~,\\
V_{40}&=&\frac{1}{4}\sum_{klij}u_ku_lv_iv_j
\overline{V}_{klij}q^{\dag}_kq^{\dag}_l
\tilde{q}^{\dag}_j\tilde{q}^{\dag}_i~,
\label{eqn:qpoperator2}
\end{eqnarray}
with $P_k=(-)^{l_k+j_k+m_k}$, $V_{ba}=(V_{ab})^\dagger$ and
$\overline{V}_{abcd}=V_{abcd}-V_{abdc}$.

The matrix elements of quasiparticle operators are related to those
of the usual one-- and two--body operators.  We show how this works
for the operator $T_{20}$; the generalization to other operators
follows automatically.   $T_{20}$ adds two quasiparticles to any
state. From $|ij\rangle=q^{\dag}_iq^{\dag}_j|-\rangle$ and
Eq.~(\ref{eqn:t20}), we write
\begin{equation}
\langle
ij|T_{20}|-\rangle~=~\sum_{kl}T_{kl}u_kv_l\langle-|q_jq_iq^{\dag}_k
\tilde{q}^{\dag}_l|-\rangle~.
\label{eqn:t20a}
\end{equation}
Since
\begin{equation}
\langle-|q_jq_iq^{\dag}_k\tilde{q}^{\dag}_l|-\rangle~=~P_l
[-\delta_{jk}\delta_{-li}+ \delta_{-lj}\delta_{ik}]~,
\label{eqn:t20b}
\end{equation}
Eq.~(\ref{eqn:t20a}) becomes
\begin{equation}
\langle ij|T_{20}|-\rangle~=~-P_i u_jv_i\langle j|T|-i\rangle+P_j
u_iv_j \langle i|T|-j\rangle~. \label{eqn:t20ac}
\end{equation}


\begin{thebibliography}{31}
\expandafter\ifx\csname
natexlab\endcsname\relax\def\natexlab#1{#1}\fi
\expandafter\ifx\csname bibnamefont\endcsname\relax
  \def\bibnamefont#1{#1}\fi
\expandafter\ifx\csname bibfnamefont\endcsname\relax
  \def\bibfnamefont#1{#1}\fi
\expandafter\ifx\csname citenamefont\endcsname\relax
  \def\citenamefont#1{#1}\fi
\expandafter\ifx\csname url\endcsname\relax
  \def\url#1{\texttt{#1}}\fi
\expandafter\ifx\csname urlprefix\endcsname\relax\def\urlprefix{URL
}\fi \providecommand{\bibinfo}[2]{#2}
\providecommand{\eprint}[2][]{\url{#2}}

\bibitem[{\citenamefont{Romalis et~al.}(2001)\citenamefont{Romalis, Griffith,
  Jacobs, and Fortson}}]{romalis01}
\bibinfo{author}{\bibfnamefont{M.~V.} \bibnamefont{Romalis}},
  \bibinfo{author}{\bibfnamefont{W.~C.} \bibnamefont{Griffith}},
  \bibinfo{author}{\bibfnamefont{J.~P.} \bibnamefont{Jacobs}},
  \bibnamefont{and} \bibinfo{author}{\bibfnamefont{E.~N.}
  \bibnamefont{Fortson}}, \bibinfo{journal}{Phys.\ Rev.\ Lett.}
  \textbf{\bibinfo{volume}{86}}, \bibinfo{pages}{2505} (\bibinfo{year}{2001}).

\bibitem[{\citenamefont{Schiff}(1963)}]{schiff63}
\bibinfo{author}{\bibfnamefont{L.~I.} \bibnamefont{Schiff}},
  \bibinfo{journal}{Phys.\ Rev.} \textbf{\bibinfo{volume}{132}},
  \bibinfo{pages}{2194} (\bibinfo{year}{1963}).

\bibitem[{\citenamefont{Herczeg}(1988)}]{herczeg87}
\bibinfo{author}{\bibfnamefont{P.}~\bibnamefont{Herczeg}},
  \emph{\bibinfo{title}{\textrm{in} Tests of Time Reversal Invariance in
  Neutron Physics}} (\bibinfo{publisher}{World Scientific},
  \bibinfo{address}{Singapore, New Jersey, Hong Kong}, \bibinfo{year}{1988}),
  \bibinfo{note}{n. R. Rogerson, C. R. Gould, and J. D. Bowman, eds., page 24}.

\bibitem[{\citenamefont{Griffiths and Vogel}(1991)}]{griffiths91}
\bibinfo{author}{\bibfnamefont{A.}~\bibnamefont{Griffiths}} \bibnamefont{and}
  \bibinfo{author}{\bibfnamefont{P.}~\bibnamefont{Vogel}},
  \bibinfo{journal}{Phys.\ Rev.\ C} \textbf{\bibinfo{volume}{44}},
  \bibinfo{pages}{1195} (\bibinfo{year}{1991}).

\bibitem[{\citenamefont{Towner and Hayes}(1994)}]{towner94}
\bibinfo{author}{\bibfnamefont{I.~S.} \bibnamefont{Towner}} \bibnamefont{and}
  \bibinfo{author}{\bibfnamefont{A.~C.} \bibnamefont{Hayes}},
  \bibinfo{journal}{Phys.\ Rev.\ C} \textbf{\bibinfo{volume}{49}},
  \bibinfo{pages}{2391} (\bibinfo{year}{1994}).

\bibitem[{\citenamefont{Flambaum et~al.}(1986)\citenamefont{Flambaum,
  Khriplovich, and Sushkov}}]{flambaum86}
\bibinfo{author}{\bibfnamefont{V.~V.} \bibnamefont{Flambaum}},
  \bibinfo{author}{\bibfnamefont{I.~B.} \bibnamefont{Khriplovich}},
  \bibnamefont{and} \bibinfo{author}{\bibfnamefont{O.~P.}
  \bibnamefont{Sushkov}}, \bibinfo{journal}{Nucl.\ Phys.}
  \textbf{\bibinfo{volume}{A449}}, \bibinfo{pages}{750} (\bibinfo{year}{1986}).

\bibitem[{\citenamefont{Flambaum and Ginges}(2002)}]{flambaum02}
\bibinfo{author}{\bibfnamefont{V.~V.} \bibnamefont{Flambaum}} \bibnamefont{and}
  \bibinfo{author}{\bibfnamefont{J.~S.~M.} \bibnamefont{Ginges}},
  \bibinfo{journal}{Phys.\ Rev.\ A} \textbf{\bibinfo{volume}{65}},
  \bibinfo{pages}{032113} (\bibinfo{year}{2002}).

\bibitem[{\citenamefont{Dmitriev and Sen'kov}(2003)}]{dmitriev03}
\bibinfo{author}{\bibfnamefont{V.~F.} \bibnamefont{Dmitriev}} \bibnamefont{and}
  \bibinfo{author}{\bibfnamefont{R.~A.} \bibnamefont{Sen'kov}},
  \bibinfo{journal}{Phys.\ Atom.\ Nucl.} \textbf{\bibinfo{volume}{66}},
  \bibinfo{pages}{1940} (\bibinfo{year}{2003}).

\bibitem[{\citenamefont{Dmitriev et~al.}(2005)\citenamefont{Dmitriev, Sen'kov,
  and Auerbach}}]{dmitriev05}
\bibinfo{author}{\bibfnamefont{V.~F.} \bibnamefont{Dmitriev}},
  \bibinfo{author}{\bibfnamefont{R.~A.} \bibnamefont{Sen'kov}},
  \bibnamefont{and} \bibinfo{author}{\bibfnamefont{N.}~\bibnamefont{Auerbach}}
  (\bibinfo{year}{2005}).

\bibitem[{\citenamefont{Brown}(1970)}]{brown70}
\bibinfo{author}{\bibfnamefont{G.~E.} \bibnamefont{Brown}},
  \emph{\bibinfo{title}{\textrm{in} Facets of Physics}}
  (\bibinfo{publisher}{Academic Press}, \bibinfo{address}{New York},
  \bibinfo{year}{1970}), \bibinfo{note}{d. Allen Bromley and Vernon W. Hughes,
  eds.}

\bibitem[{\citenamefont{Ellis and Osnes}(1997)}]{ellis97}
\bibinfo{author}{\bibfnamefont{P.~J.} \bibnamefont{Ellis}} \bibnamefont{and}
  \bibinfo{author}{\bibfnamefont{E.}~\bibnamefont{Osnes}},
  \bibinfo{journal}{Rev.\ Mod.\ Phys.} \textbf{\bibinfo{volume}{1997}},
  \bibinfo{pages}{777} (\bibinfo{year}{1997}).

\bibitem[{\citenamefont{Bender et~al.}(2002)\citenamefont{Bender, Dobaczewski,
  Engel, and Nazarewicz}}]{bender02}
\bibinfo{author}{\bibfnamefont{M.}~\bibnamefont{Bender}},
  \bibinfo{author}{\bibfnamefont{J.}~\bibnamefont{Dobaczewski}},
  \bibinfo{author}{\bibfnamefont{J.}~\bibnamefont{Engel}}, \bibnamefont{and}
  \bibinfo{author}{\bibfnamefont{W.}~\bibnamefont{Nazarewicz}},
  \bibinfo{journal}{Phys.\ Rev.\ C} \textbf{\bibinfo{volume}{65}},
  \bibinfo{pages}{054322} (\bibinfo{year}{2002}).

\bibitem[{\citenamefont{Sushkov et~al.}(1984)\citenamefont{Sushkov, Flambaum,
  and Khriplovich}}]{sushkov84}
\bibinfo{author}{\bibfnamefont{O.~P.} \bibnamefont{Sushkov}},
  \bibinfo{author}{\bibfnamefont{V.~V.} \bibnamefont{Flambaum}},
  \bibnamefont{and} \bibinfo{author}{\bibfnamefont{I.~B.}
  \bibnamefont{Khriplovich}}, \bibinfo{journal}{Sov. Phys. JETP}
  \textbf{\bibinfo{volume}{60}}, \bibinfo{pages}{873} (\bibinfo{year}{1984}).

\bibitem[{\citenamefont{Terasaki et~al.}(2005)\citenamefont{Terasaki, Engel,
  Bender, Dobaczewski, Nazarewicz, and Stoitsov}}]{terasaki05}
\bibinfo{author}{\bibfnamefont{J.}~\bibnamefont{Terasaki}},
  \bibinfo{author}{\bibfnamefont{J.}~\bibnamefont{Engel}},
  \bibinfo{author}{\bibfnamefont{M.}~\bibnamefont{Bender}},
  \bibinfo{author}{\bibfnamefont{J.}~\bibnamefont{Dobaczewski}},
  \bibinfo{author}{\bibfnamefont{W.}~\bibnamefont{Nazarewicz}},
  \bibnamefont{and} \bibinfo{author}{\bibfnamefont{M.}~\bibnamefont{Stoitsov}},
  \bibinfo{journal}{Phys.\ Rev.\ C} \textbf{\bibinfo{volume}{71}},
  \bibinfo{pages}{034310} (\bibinfo{year}{2005}).

\bibitem[{\citenamefont{Agrawal et~al.}(2003)\citenamefont{Agrawal, Shlomo, and
  Sanzhur}}]{agrawal03}
\bibinfo{author}{\bibfnamefont{B.}~\bibnamefont{Agrawal}},
  \bibinfo{author}{\bibfnamefont{S.}~\bibnamefont{Shlomo}}, \bibnamefont{and}
  \bibinfo{author}{\bibfnamefont{A.}~\bibnamefont{Sanzhur}},
  \bibinfo{journal}{Phys.\ Rev.\ C} \textbf{\bibinfo{volume}{67}},
  \bibinfo{pages}{034314} (\bibinfo{year}{2003}).

\bibitem[{\citenamefont{Reinhard et~al.}(1999)\citenamefont{Reinhard, Dean,
  Nazarewicz, Dobaczewski, Maruhn, and Strayer}}]{reinhard99}
\bibinfo{author}{\bibfnamefont{P.-G.} \bibnamefont{Reinhard}},
  \bibinfo{author}{\bibfnamefont{D.~J.} \bibnamefont{Dean}},
  \bibinfo{author}{\bibfnamefont{W.}~\bibnamefont{Nazarewicz}},
  \bibinfo{author}{\bibfnamefont{J.}~\bibnamefont{Dobaczewski}},
  \bibinfo{author}{\bibfnamefont{J.~A.} \bibnamefont{Maruhn}},
  \bibnamefont{and} \bibinfo{author}{\bibfnamefont{M.~R.}
  \bibnamefont{Strayer}}, \bibinfo{journal}{Phys.\ Rev.\ C}
  \textbf{\bibinfo{volume}{60}}, \bibinfo{pages}{014316}
  (\bibinfo{year}{1999}).

\bibitem[{\citenamefont{Engel et~al.}(2003)\citenamefont{Engel, Bender,
  Dobaczewski, de~Jesus, and Olbratowski}}]{engel03}
\bibinfo{author}{\bibfnamefont{J.}~\bibnamefont{Engel}},
  \bibinfo{author}{\bibfnamefont{M.}~\bibnamefont{Bender}},
  \bibinfo{author}{\bibfnamefont{J.}~\bibnamefont{Dobaczewski}},
  \bibinfo{author}{\bibfnamefont{J.~H.} \bibnamefont{de~Jesus}},
  \bibnamefont{and}
  \bibinfo{author}{\bibfnamefont{P.}~\bibnamefont{Olbratowski}},
  \bibinfo{journal}{Phys.\ Rev.\ C} \textbf{\bibinfo{volume}{68}},
  \bibinfo{pages}{025501} (\bibinfo{year}{2003}).

\bibitem[{\citenamefont{Beiner et~al.}(1975)\citenamefont{Beiner, Flocard,
  Giai, and Quentin}}]{beiner75}
\bibinfo{author}{\bibfnamefont{M.}~\bibnamefont{Beiner}},
  \bibinfo{author}{\bibfnamefont{H.}~\bibnamefont{Flocard}},
  \bibinfo{author}{\bibfnamefont{N.~V.} \bibnamefont{Giai}}, \bibnamefont{and}
  \bibinfo{author}{\bibfnamefont{P.}~\bibnamefont{Quentin}},
  \bibinfo{journal}{Nucl.\ Phys.} \textbf{\bibinfo{volume}{A238}},
  \bibinfo{pages}{29} (\bibinfo{year}{1975}).

\bibitem[{\citenamefont{Bartel et~al.}(1982)\citenamefont{Bartel, Quentin,
  Brack, Guet, and H{\aa}kansson}}]{bartel82}
\bibinfo{author}{\bibfnamefont{J.}~\bibnamefont{Bartel}},
  \bibinfo{author}{\bibfnamefont{P.}~\bibnamefont{Quentin}},
  \bibinfo{author}{\bibfnamefont{M.}~\bibnamefont{Brack}},
  \bibinfo{author}{\bibfnamefont{C.}~\bibnamefont{Guet}}, \bibnamefont{and}
  \bibinfo{author}{\bibfnamefont{H.~B.} \bibnamefont{H{\aa}kansson}},
  \bibinfo{journal}{Nucl.\ Phys.} \textbf{\bibinfo{volume}{A386}},
  \bibinfo{pages}{79} (\bibinfo{year}{1982}).

\bibitem[{\citenamefont{Chabanat et~al.}(1998)\citenamefont{Chabanat, Bonche,
  Haensel, Meyer, and Schaeffer}}]{chabanat98}
\bibinfo{author}{\bibfnamefont{E.}~\bibnamefont{Chabanat}},
  \bibinfo{author}{\bibfnamefont{P.}~\bibnamefont{Bonche}},
  \bibinfo{author}{\bibfnamefont{P.}~\bibnamefont{Haensel}},
  \bibinfo{author}{\bibfnamefont{J.}~\bibnamefont{Meyer}}, \bibnamefont{and}
  \bibinfo{author}{\bibfnamefont{R.}~\bibnamefont{Schaeffer}},
  \bibinfo{journal}{Nucl.\ Phys.} \textbf{\bibinfo{volume}{A635}},
  \bibinfo{pages}{231} (\bibinfo{year}{1998}), \bibinfo{note}{{N}ucl.\ Phys.\
  \textbf{A643}, 441(E) (1998).}

\bibitem[{\citenamefont{Dobaczewski et~al.}(1984)\citenamefont{Dobaczewski,
  Flocard, and Treiner}}]{dobaczewski84}
\bibinfo{author}{\bibfnamefont{J.}~\bibnamefont{Dobaczewski}},
  \bibinfo{author}{\bibfnamefont{H.}~\bibnamefont{Flocard}}, \bibnamefont{and}
  \bibinfo{author}{\bibfnamefont{J.}~\bibnamefont{Treiner}},
  \bibinfo{journal}{Nucl.\ Phys.\ \textbf{A}} \textbf{\bibinfo{volume}{422}},
  \bibinfo{pages}{103} (\bibinfo{year}{1984}).

\bibitem[{\citenamefont{Clark et~al.}(2001)\citenamefont{Clark, Lui, and
  Youngblood}}]{clark01}
\bibinfo{author}{\bibfnamefont{H.~L.} \bibnamefont{Clark}},
  \bibinfo{author}{\bibfnamefont{Y.-W.} \bibnamefont{Lui}}, \bibnamefont{and}
  \bibinfo{author}{\bibfnamefont{D.~H.} \bibnamefont{Youngblood}},
  \bibinfo{journal}{Phys. Rev. C} \textbf{\bibinfo{volume}{63}},
  \bibinfo{pages}{031301(R)} (\bibinfo{year}{2001}).

\bibitem[{\citenamefont{Davis et~al.}(1997)}]{davis97}
\bibinfo{author}{\bibfnamefont{B.~F.} \bibnamefont{Davis}}
  \bibnamefont{et~al.}, \bibinfo{journal}{Phys. Rev. Lett.}
  \textbf{\bibinfo{volume}{79}}, \bibinfo{pages}{609} (\bibinfo{year}{1997}).

\bibitem[{\citenamefont{Hamamoto et~al.}(1998)\citenamefont{Hamamoto, Sagawa,
  and Zhang}}]{hamamoto98}
\bibinfo{author}{\bibfnamefont{I.}~\bibnamefont{Hamamoto}},
  \bibinfo{author}{\bibfnamefont{H.}~\bibnamefont{Sagawa}}, \bibnamefont{and}
  \bibinfo{author}{\bibfnamefont{X.~Z.} \bibnamefont{Zhang}},
  \bibinfo{journal}{Phys. Rev. C} \textbf{\bibinfo{volume}{57}},
  \bibinfo{pages}{R1064} (\bibinfo{year}{1998}).

\bibitem[{\citenamefont{Col\`{o} et~al.}(2000)\citenamefont{Col\`{o}, Giai,
  Bortignon, and Quaglia}}]{colo00}
\bibinfo{author}{\bibfnamefont{G.}~\bibnamefont{Col\`{o}}},
  \bibinfo{author}{\bibfnamefont{N.~V.} \bibnamefont{Giai}},
  \bibinfo{author}{\bibfnamefont{P.~F.} \bibnamefont{Bortignon}},
  \bibnamefont{and} \bibinfo{author}{\bibfnamefont{M.~R.}
  \bibnamefont{Quaglia}}, \bibinfo{journal}{Phys. Lett.}
  \textbf{\bibinfo{volume}{B485}}, \bibinfo{pages}{362} (\bibinfo{year}{2000}).

\bibitem[{\citenamefont{Vretenar et~al.}(2000)\citenamefont{Vretenar, Wamdelt,
  and Ring}}]{vretenar00}
\bibinfo{author}{\bibfnamefont{D.}~\bibnamefont{Vretenar}},
  \bibinfo{author}{\bibfnamefont{A.}~\bibnamefont{Wamdelt}}, \bibnamefont{and}
  \bibinfo{author}{\bibfnamefont{P.}~\bibnamefont{Ring}},
  \bibinfo{journal}{Phys. Lett.} \textbf{\bibinfo{volume}{B487}},
  \bibinfo{pages}{334} (\bibinfo{year}{2000}).

\bibitem[{\citenamefont{Abrosimov et~al.}(2002)\citenamefont{Abrosimov,
  Dellafiore, and Matera}}]{abrosimov02}
\bibinfo{author}{\bibfnamefont{V.~I.} \bibnamefont{Abrosimov}},
  \bibinfo{author}{\bibfnamefont{A.}~\bibnamefont{Dellafiore}},
  \bibnamefont{and} \bibinfo{author}{\bibfnamefont{F.}~\bibnamefont{Matera}},
  \bibinfo{journal}{Nucl. Phys.} \textbf{\bibinfo{volume}{A697}},
  \bibinfo{pages}{748} (\bibinfo{year}{2002}).

\bibitem[{\citenamefont{Shlomo and Sanzhur}(2002)}]{shlomo02}
\bibinfo{author}{\bibfnamefont{S.}~\bibnamefont{Shlomo}} \bibnamefont{and}
  \bibinfo{author}{\bibfnamefont{A.}~\bibnamefont{Sanzhur}},
  \bibinfo{journal}{Phys. Rev. C} \textbf{\bibinfo{volume}{65}},
  \bibinfo{pages}{044310} (\bibinfo{year}{2002}).

\bibitem[{\citenamefont{Engel et~al.}(1999)\citenamefont{Engel, Friar, and
  Hayes}}]{engel99a}
\bibinfo{author}{\bibfnamefont{J.}~\bibnamefont{Engel}},
  \bibinfo{author}{\bibfnamefont{J.~L.} \bibnamefont{Friar}}, \bibnamefont{and}
  \bibinfo{author}{\bibfnamefont{A.~C.} \bibnamefont{Hayes}},
  \bibinfo{journal}{Phys.\ Rev.\ C} \textbf{\bibinfo{volume}{61}},
  \bibinfo{pages}{035502} (\bibinfo{year}{1999}).

\bibitem[{\citenamefont{Dobaczewski and Engel}(2005)}]{dobaczewski05}
\bibinfo{author}{\bibfnamefont{J.}~\bibnamefont{Dobaczewski}} \bibnamefont{and}
  \bibinfo{author}{\bibfnamefont{J.}~\bibnamefont{Engel}},
  \bibinfo{journal}{Submmitted to Phys. Rev. Lett.}  (\bibinfo{year}{2005}).

\bibitem[{\citenamefont{Dzuba et~al.}(2002)\citenamefont{Dzuba, Flambaum,
  Ginges, and Kozlov}}]{dzuba02}
\bibinfo{author}{\bibfnamefont{V.~A.} \bibnamefont{Dzuba}},
  \bibinfo{author}{\bibfnamefont{V.~V.} \bibnamefont{Flambaum}},
  \bibinfo{author}{\bibfnamefont{J.~S.~M.} \bibnamefont{Ginges}},
  \bibnamefont{and} \bibinfo{author}{\bibfnamefont{M.~G.}
  \bibnamefont{Kozlov}}, \bibinfo{journal}{Phys.\ Phys.\ A}
  \textbf{\bibinfo{volume}{66}}, \bibinfo{pages}{012111}
  (\bibinfo{year}{2002}).

\end{thebibliography}
\end{document}